\documentclass[10pt,aps,pra,superscriptaddress,twocolumn,showpacs,preprintnumbers,
nofootinbib]{revtex4-1}
\usepackage{amsfonts}
\usepackage{amssymb,amsmath}
\usepackage{mathrsfs}
\usepackage{amsthm}
\usepackage{bm}
\usepackage{newlfont}
\usepackage{graphicx}

\newcommand{\usb}{\affiliation{Departamento de F\'{\i}sica, Secci\'{o}n de Fen\'{o}menos \'{O}pticos, 
Universidad Sim\'{o}n Bol\'{\i}var,Apartado Postal 89000, Caracas 1080-A, Venezuela.}}
\newcommand{\ivic}{\affiliation{Centro de F\'{\i}sica, Instituto Venezolano de Investigaciones 
Cient\'{\i}ficas, Apartado 20632 Caracas 1020-A, Venezuela.}}

\begin{document}
\begin{flushright} ${}$\\[-40pt] $\scriptstyle \mathrm SB/F/xxx-16$ \\[0pt]
\end{flushright}
\title{An elementary approach to electromagnetic momentum in matter} 
\author{Rodrigo Medina}\email[]{rmedina@ivic.gob.ve}\ivic
\author{J Stephany}\email[]{stephany@usb.ve}\usb

\begin{abstract}
We present an elementary discussion of the momentum transferred to a conducting 
sheet by an electromagnetic wave propagating in a polarizable medium. 
We show that conservation of momentum is consistent with Minkowski's 
expression for the momentum density.
\end{abstract}
\maketitle 
\thispagestyle{empty} 
\newpage
\section{Introduction}
Although the conservation of energy and momentum of electromagnetic fields in
vacuum is a very well  understood problem,  inside a polarizable medium it is
still a matter of discussion. A review of the literature on this subject can be
found in \cite{Griffiths2012}. 

In the vacuum, Poynting's equation for the exchange of energy between the field and
free charges is easily derived starting from  Lorentz's force density on charges 
and currents
\begin{equation}
\label{force-density}
\mathbf{f}=\rho \mathbf{E}+\mathbf{j}\times\mathbf{B}
\end{equation}
and from the power density on the currents
\begin{equation}
\label{power-density}
\dot{w}=\mathbf{j}\cdot\mathbf{E}\ .
\end{equation}
 The resulting equation
\begin{equation}
\label{Poynting-eq}
\frac{\partial u}{\partial t} + \nabla\cdot\mathbf{S}=
-\mathbf{j}\cdot\mathbf{E}
\end{equation}
written in terms of 
\begin{equation}
\label{energy-density}
u=u_\mathrm{v}=\frac{1}{2}\epsilon_0 E^2 + \frac{1}{2\mu_0}B^2
\end{equation}
and
\begin{equation}
\mathbf{S}=\frac{1}{\mu_0}\mathbf{E}\times\mathbf{B}
\end{equation}
is an identity valid for any solution of Maxwell's equations. It is
interpreted as a local energy conservation equation by taking $u_\mathrm{v}$ as 
the energy density of the electromagnetic field and Poynting's vector
$\mathbf{S}$ as the energy flux current vector. The power exerted by the matter
on the fields is assumed to be the opposite of (\ref{power-density}).

Electromagnetic fields also carry momentum. The derivation of the equation for the momentum 
exchange requires the use of Maxwell's stress  tensor  and a somewhat more involved calculation
which do not make easy an elementary presentation of the subject. In many  textbooks  
(e.g \cite{Halliday2014}, \cite{Sears2012}) radiation pressure is simply 
introduced as a new fact, sometimes referring to it as one of Maxwell's achievements. In the vacuum 
the momentum density is accepted to be
\begin{equation}
\label{momentum-density}
\mathbf{g}=\epsilon_0\mathbf{E}\times\mathbf{B}=\frac{1}{c^2}\mathbf{S}\ .
\end{equation}
which leads to the relation 
\begin{equation}
\label{Deltap}
 \Delta p = \frac{\Delta U}{c}
\end{equation} 
for the energy and momentum that an electromagnetic wave releases when absorbed.
Because of this in other 
elementary presentations (e.g \cite{Alonso1992}) the density of momentum 
(\ref{momentum-density}) is justified  by working on the analogy with a relativistic 
particle, whose momentum and energy are related by 
\begin{equation}
\label{particle-momentum} 
\mathbf{p}=\frac{E}{c^2}\mathbf{v}\ .
\end{equation}
For an ultra relativistic particle ($v \to c$) holds
\begin{equation}
\label{particle-momentum-ultra}
\mathbf{p}=\frac{E}{c}\hat{\mathbf{v}}\ .
\end{equation}
By assuming that this last relation is also valid for a packet of
electromagnetic waves traveling in some direction, one gets (\ref{Deltap})
and therefore $g=c^{-2}S$ in accord with (\ref{momentum-density}). This procedure
has the shortcoming of having a non electromagnetic element, the
relativistic particle, introduced in the discussion.

A more enlighting treatment of the subject is the one presented  for example in 
Ref.\cite{Resnik}. There, the interaction of 
electromagnetic wave packets coming from vacuum with electric currents in a 
non ideal conductor are considered. The impulse transferred to the currents
is computed and showed to be consistent with (\ref{momentum-density}) and (\ref{Deltap}). 
In this paper we show that this same idea can  also be applied to the propagation of 
light in matter.

As we already commented, what happens with momentum inside matter is still not 
completely clear from the theoretical point of view. For
the energy, Poynting \cite{Poynting} proposed that Eq.\,(\ref{Poynting-eq}) 
remains valid with
\begin{equation}
\label{Poynting-density}
u=\frac{1}{2}\mathbf{D}\cdot\mathbf{E}+\frac{1}{2}\mathbf{H}\cdot\mathbf{B}
\end{equation} 
and
\begin{equation}
\label{Poynting-vector}
\mathbf{S}=\mathbf{E}\times\mathbf{H}\ .
\end{equation}
Here, $\mathbf{j}$ is the free current density,
 $\mathbf{D}=\epsilon_0\mathbf{E}+\mathbf{P}$ is the electric displacement
and $\mathbf{H}=1/\mu_0\mathbf{B}-\mathbf{M}$ is the magnetizing field.
It may be shown that Eq.(\ref{Poynting-eq}) is consistent with 
Maxwell's equations only for materials with linear polarizations, $\mathbf{P}\propto\mathbf{E}$ and
 $\mathbf{M}\propto\mathbf{B}$. In particular it holds for a material with
permittivity $\epsilon$ and permeability $\mu$,
 $\mathbf{P}=(\epsilon-\epsilon_0)\mathbf{E}$
and  $\mathbf{M}=[(\mu-\mu_0)/\mu_0\mu]\mathbf{B}$, but it also holds in the more general
situation of a non-isotropic material.

Poynting's density is the vacuum energy density $u_\mathrm{v}$ plus
terms that  depend on  the polarization $\mathbf{P}$ and the magnetization
$\mathbf{M}$,
\begin{equation}
u=u_\mathrm{v}+\frac{1}{2}\mathbf{P}\cdot\mathbf{E}
-\frac{1}{2}\mathbf{M}\cdot\mathbf{B} \ .
\end{equation}
The electric term corresponds to the energy density of deformation of matter
as it develops electric dipole moments. The magnetic term is the work density 
made on microscopic currents by the induced electric field. The work done on
electric and magnetic moments is already included in Poynting's energy
density, that is the reason why in the right hand side of equation
(\ref{Poynting-eq}) only the power of the free current density appears. In
other words, for an electromagnetic wave in a material medium, Poynting's 
expression corresponds to  the energy density of the whole wave, including 
the change of energy of matter as the wave propagates in the medium. 

For the momentum density of the fields in matter the situation is less clear.
There is a dispute that has lasted more than a century \cite{Griffiths2012} 
on the subject. Minkowski \cite{ Minkowski1908} proposed
\begin{equation}
\label{Minkowski}
\mathbf{g}_{\mathrm{Min}}=\mathbf{D}\times\mathbf{B}\ ,
\end{equation}
which for individual photons corresponds to using (\ref{particle-momentum-ultra}) replacing $c$ by
the speed of light  in matter $v=(\epsilon\mu)^{-1/2}$. This was challenged by Abraham \cite{Abraham1909,Abraham1910} who
proposed instead 
\begin{equation}
\label{Abraham}
\mathbf{g}_{\mathrm{Abr}}=\frac{1}{c^2}\mathbf{E}\times\mathbf{H}=\frac{1}{c^2}\mathbf{S}\ ,
\end{equation}
which for photons corresponds to using directly (\ref{particle-momentum}).

\begin{figure}
\centering
\includegraphics[scale=0.7]{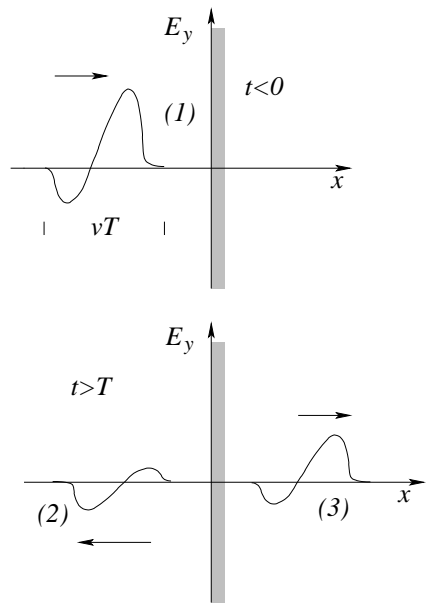}
\caption{Incident (1), reflected (2) and transmitted (3) packets}
\label{paquetes}
\end{figure}

To shed some light on this issue, in this paper we study a simple example that can be solved exactly. It
consists of a packet of plane waves propagating in a material with permeability
$\mu$ and permittivity $\epsilon$. The medium being homogeneous there is
no refraction and the wave travels conserving momentum and energy. Inside the
material there is a flat thin film of conducting material in which Ohm's law
holds, $\mathbf{j}\propto \mathbf{E}$. When the wave packet falls on the film
it splits in  reflected and  transmitted packets, whose amplitudes are 
fixed by the boundary conditions (See Figure [\ref{paquetes}]). The total work done on the
currents in the conducting film can be calculated using (\ref{power-density}). The result
is consistent with Poynting's energy density (\ref{Poynting-density}). Also, the impulse
on the currents of the film can be calculated using (\ref{force-density}). We found
that the result is consistent with momentum conservation  only if the momentum carried by the waves
is given by Minkowski's expression (\ref{Minkowski}) .
\section{Momentum transfer}
\label{Momentum transfer}

Consider a medium with electric and magnetic permeabilities $\epsilon$ and $\mu$ filling the space. 
At the plane $x=0$ there is a sheet of thickness $a$ made of a conductive material of resistivity $\rho$.
A packet of electromagnetic plane waves polarized in the direction $y$ is traveling in the $x$ direction.
Suppose it has a spatial length $vT$ with $v=1/\sqrt{\epsilon\mu}$, $T$ a time scale, and a wave length 
$\lambda$ such that $a<< \lambda$ and $a<<vT$. In terms of a suitable function $h(t)$ which vanishes outside 
the interval $[0,T]$ the incident wave is,
\begin{eqnarray}
\mathbf{E}_1(x,y,z,t)&=&E_1 h(t-x/v)\theta(-x)\hat{\mathbf{y}} \ \ \ \ x<0\\
\mathbf{B}_1(x,y,z,t) &=& \frac{1}{v}E_1 h(t-x/v)\theta(-x)\hat{\mathbf{z}}\ .
\end{eqnarray}
The reflected and transmitted waves are,
\begin{eqnarray}
\mathbf{E}_2(x,y,z,t) &=& E_2 h(t+x/v)\theta(-x)\hat{\mathbf{y}} \ ,\\
\mathbf{B}_2(x,y,z,t) &=& -\frac{1}{v}E_2h(t+x/v)\theta(-x)\hat{\mathbf{z}}\ ,
\end{eqnarray}
and
\begin{eqnarray}
\mathbf{E}_3(x,y,z,t) &=& E_3 h(t-x/v)\theta(x)\hat{\mathbf{y}} \ ,\\
\mathbf{B}_3(x,y,z,t) &=& \frac{1}{v}E_3 h(t-x/v)\theta(x)\hat{\mathbf{z}}\ .
\end{eqnarray}
At $x=0$ $\mathbf{E}$ is continuous, $\mathbf{E}(0^+,y,z,t)=\mathbf{E}(0^-,y,z,t)$ implying
\begin{equation}
 E_1+E_2=E_3\ .
\end{equation}

\begin{figure}
\centering
\includegraphics[scale=0.8]{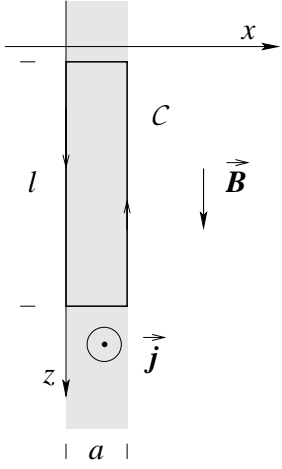}
\caption{The integration path}
\label{camino}
\end{figure}

There is an electrical current in the sheet. The magnetic field is discontinuous at $x=0$.
Taking a rectangular path of length $l$ in the $z$ direction and  width $a$ in the $x$ direction we 
have,
\begin{equation}
 \oint_C \mathbf{H}\cdot d\mathbf{l}=\int_S \mathbf{j}\cdot d\mathbf{S}+
 \frac{d}{dt}\int_S \mathbf{D}\cdot d\mathbf{S}
\end{equation}
where $S$ is the plane surface with boundary $C$ (See Figure [\ref{camino}]). The second integral vanishes 
because the width of the surface is negligible. The current density is
\begin{equation}
\label{currentdensity}
 \mathbf{j}=\frac{1}{\rho}\mathbf{E}=\frac{1}{\rho}E_3h(t)\hat{\mathbf{y}}\ .
\end{equation}
Substituting this in the previous equation we have
\begin{equation}
\frac{1}{\mu}(B_1-B_2-B_3)h(t)l= j a l
\end{equation}
and
\begin{equation}
\frac{1}{v}(E_1-E_2-E_3)=\frac{\mu a E_3}{\rho}\ .
\end{equation}
Introducing
\begin{equation}
  b=\frac{ \mu a v}{2\rho}
\end{equation}
and solving the system we have,
\begin{eqnarray}
\label{solution}
E_2=-\frac{b E_1}{1+b}\ \ \ ,\ \ \ E_3=\frac{E_1}{1+b}\ .
\end{eqnarray}
The energy of the wave is given by Poynting's expression
\begin{equation}
 u=\frac{1}{2}(\epsilon E^2+\frac{1}{\mu}B^2)=\epsilon E^2\ .
\end{equation}
For $t<0$ we consider a cylindrical piece of
the incident packet with axis parallel to $x$ and cross section  $A$. The energy is,
\begin{eqnarray}
 u&=&\epsilon E_1^2h^2(t-x/v)\ ,\\
 U_1&=&\int \epsilon E_1^2h^2(t-x/v)d^3r=\epsilon E_1^2 A v \bar{T}
\end{eqnarray}
where 
\begin{equation}
 \bar{T}=\int_0^Th^2(t) dt\ .
\end{equation}
For $t>T$ the energies of the  reflected and transmitted  waves are,
\begin{eqnarray}
 U_2&=&\epsilon E_2^2 A v \bar{T}=\left[\frac{b}{1+b}\right]^2U_1\ ,\\
 U_3&=&\epsilon E_3^2 A v \bar{T}=\left[\frac{1}{1+b}\right]^2U_1\ .
\end{eqnarray}
The power density on the current is
\begin{equation}
\dot{w}=\mathbf{j}\cdot \mathbf{E}=\frac{E^2(t)}{\rho}=\frac{E_3^2h^2(t)}{\rho}\ .
\end{equation}
The total power transferred by the wave is,
\begin{equation}
\frac{dW}{dt}=\int  \mathbf{E}\cdot \mathbf{j}\, d^3r = \frac{E_3^2Aa h(t)^2}{\rho}
\end{equation}
and the work on the current is
\begin{equation}
 W=\int \frac{dW}{dt}dt= \frac{E_3^2Aa \bar{T}}{\rho}=\frac{2b}{[1+b]^2}U_1\ .
\end{equation} 
The energy is conserved
\begin{equation}
 U_1=U_2+U_3+W\ .
\end{equation}
Let us now discuss momentum conservation for this system. We compute first 
the force applied on the sheet. The force density on the current is
\begin{equation}
 \mathbf{f}=\mathbf{j} \times \mathbf{B}\ .
\end{equation}
In the approximation where $a$ is infinitesimal, the current density 
in the sheet, $ \mathbf{j}$, is
constant because the electric field is continuous. 
The magnetic field then varies linearly between its values at both sides of the sheet. In 
computing the total force one has to take the average
on the sheet volume. For $0\leq t\leq T$ we have
\begin{eqnarray}
\label{force}
 \mathbf{F} &=& \int \mathbf{f}  d^3r = jA\int_0^a B_z(x)dx \hat{\mathbf{x}}\nonumber\\
&=&jAa\frac{(B_z(0^{-})+B_z(0^{+}))}{2} \hat{\mathbf{x}}
\end{eqnarray}
where
\begin{eqnarray}
\label{averageB}
 \frac{1}{2}(B_z(0^-)+B_z(0^+))&=&\frac{h(t)}{2v}[E_1-E_2+E_3]\nonumber\\
 &=&\frac{h(t)}{v}E_1\ .
\end{eqnarray}
Substituting this in (\ref{force}) and using (\ref{currentdensity}) and (\ref{solution}) we get
\begin{equation}
 \mathbf{F}=\frac{Aah^2(t)E_1^2}{v\rho( 1+b)}\hat{\mathbf{x}}\ .
\end{equation}
The impulse applied to the current sheet is
\begin{eqnarray}
\label{applImp}
 \mathbf{I}&=&\int \mathbf{F} dt=\frac{Aa\bar{T}E_1^2}{v\rho( 1+b)}\hat{\mathbf{x}}\nonumber\\
 &=&\frac{2b}{(1+b)v}U_1\hat{\mathbf{x}}\ .
\end{eqnarray}

The momentum carried by the wave is directed in the same direction in which the wave propagates, 
that is,  the direction of $ \mathbf{E}\times \mathbf{B}$. The momentum density of the wave 
may be written in the form,
\begin{equation}
\label{momden}
  \mathbf{g}={\alpha} \mathbf{E}\times\mathbf{B}\ ,
\end{equation}
with $\alpha$ a constant to be determined. For $\alpha=\epsilon$ the momentum density 
is Minkowski's and for $\alpha=1/c^2\mu$ it is Abraham's. Then, the momentum carried by the incident wave is
\begin{eqnarray}
  \mathbf{p}_1&=&\int\mathbf{g} _1 dV=\frac{\alpha E_1^2}{v}\int h^2(t-x/v)\hat{\mathbf{x}}d^3r\\
   &=&\alpha E_1^2A\bar{T}\hat{\mathbf{x}}=\frac{\alpha U_1}{\epsilon v}\hat{\mathbf{x}}\ .
\end{eqnarray}
Analogously
\begin{equation}
 \mathbf{p}_2=-\frac{\alpha U_2}{\epsilon v}\hat{\mathbf{x}}\ \ \ ,\ \ \ 
 \mathbf{p}_3=\frac{\alpha U_3}{\epsilon v}\hat{\mathbf{x}}\ .
\end{equation}
For the balance equation we get,
\begin{equation}
 \mathbf{p}_1-\mathbf{p}_2- \mathbf{p}_3=\frac{\alpha }{\epsilon v}(U_1+U_2-U_3)\hat{\mathbf{x}}
 =\frac{\alpha }{\epsilon } \mathbf{I}\ .
\end{equation}
Momentum is conserved if $\alpha=\epsilon$. The momentum density (\ref{momden}) turns out to be Minkowski's expression,
\begin{equation}
\label{momMin}
\mathbf{g}=\mathbf{g}_{\mathrm{Min}}= \mathbf{D}\times\mathbf{B}\ .
\end{equation}

The key in the previous argument is the factor $v^{-1}$ that appears in the
impulse (\ref{applImp}). This can be traced back to the magnetic force in
Lorentz's force (\ref{force-density}) and the relationship between the magnetic and 
electric amplitudes in electromagentic waves. This relation is a consequence of Maxwell's equations.
For obtaining Abraham's  result the factor should have been $c^{-2}v$.

\section{Conclusion}

The calculation presented in this paper assumes the validity of Maxwell's equations and 
Lorentz's force density on the conduction current.  Using the properties of the electromagnetic wave solutions of
Maxwell's equations in a simple setup we showed that momentum conservation requires the use of Minkowski's 
expression for the momentum density of the electromagnetic field in a linear medium.

Although the use of the explicit solution of Maxwell equations with boundary conditions puts our approach 
a little ahead of most introductory physics university courses it is still a very simple and elementary construction
which allows the discussion of electromagnetic momentum in material media in a self-consistent way. In particular 
it also may be used to treat the case when the wave propagates in the vacuum.



\begin{thebibliography}{long}
\bibitem{Griffiths2012} D.~J.~Griffiths, Am. J. Phys. 80, 7 (2012).
\bibitem{Halliday2014}D.~Halliday, R.~Resnik and J.~Walker,{\it Fundamentals of Physics}, 
10th Edition, John Wiley and Sons, New York, 2014, p. 983.
\bibitem{Sears2012}H.~D.~Young and R.~A.~Freedman, {\it Sears and Zemansky's University
Physics}, 13th Edition, Addison-Wesley, Boston, 2012, p. 1068.
\bibitem{Alonso1992}M.~Alonso and E.~J.~Finn, {\it Physics}, Addison-Wesley, New York, 1992, p. 788.
\bibitem{Resnik}D.~Halliday and R.~Resnik,{\it Physics, Part two}, 3d Edition, John Wiley and Sons, New York, 1978, p. 921.
\bibitem{Poynting}J.~H.~Poynting, Phil.~Trans.~R.~Soc.~ \textbf{175}, 343--361 (1884).
\bibitem{Minkowski1908} H.~Minkowski, Nachr.~Ges.~Wiss.~G\"ottingen, 53, (1908).
\bibitem{Abraham1909} M.~Abraham, Rend.~Circ.~Mat.~Palermo \textbf{28}, 1--28, (1909).
\bibitem{Abraham1910} M.~Abraham, Rend.~Circ.~Mat.~Palermo \textbf{30}, 33i--46, (1910).
\end{thebibliography}
\end{document}